\documentclass[12pt,a4paper,superscriptaddress,floatfix]{article}

\advance\voffset by -2.0cm \advance\hoffset by -1.25cm
\usepackage{amsmath}
\usepackage{amssymb}
\usepackage{amsthm}
\usepackage{amsfonts}

\usepackage{mathrsfs}

\numberwithin{equation}{section}

\theoremstyle{definition}

\newcommand{\CC}{\mathbb{C}} 
\newcommand{\RR}{\mathbb{R}} 
\newcommand{\NN}{\mathbb{N}} 

\newcommand{\be}{\begin{equation}}
\newcommand{\ee}{\end{equation}}

\def\1{\frak 1}
\def\2{\frak 2}
\def\3{\frak 3}

\newlength{\oldcolsep}

\begin{document}

\title{Quantum Field Perturbation Theory Revisited}
\author{Marco Matone}\date{}

\maketitle

\begin{center} Dipartimento di Fisica e Astronomia ``G. Galilei'' \\
 Istituto
Nazionale di Fisica Nucleare \\
Universit\`a di Padova, Via Marzolo, 8-35131 Padova,
Italy\end{center}

\bigskip

\begin{abstract}
\noindent
Schwinger's formalism in quantum field theory can be easily implemented in the case of scalar theories in $D$ dimension
with exponential interactions, such as $\mu^D\exp(\alpha\phi)$. In particular, we use the relation
 $$
 \exp\big(\alpha{\delta\over \delta J(x)}\big)\exp(-Z_0[J])=\exp(-Z_0[J+\alpha_x])
 $$
 with $J$  the external source, and $\alpha_x(y)=\alpha\delta(y-x)$.
Such a shift is strictly related to the normal ordering of $\exp(\alpha\phi)$ and to a scaling relation
which follows by renormalizing $\mu$.
Next, we derive a new formulation of perturbation theory for the potentials $V(\phi)={\lambda\over n!}:\phi^n:$, using the generating functional associated to $:\exp(\alpha\phi):$.
The $\Delta(0)$-terms related to the normal ordering are absorbed at once. The functional derivatives with respect to $J$ to compute
the generating functional are replaced by ordinary derivatives with respect to auxiliary parameters.
We focus on scalar theories, but the method is general and similar investigations extend to other theories.

\end{abstract}

\newpage

\section{Introduction}

The difficulties in quantizing some nonrenormalizable field theories, are due to the absence
of uniqueness rather than in the existence of a solution. In this respect, one should recall that
the classification of super-renormalizable, renormalizable, and nonrenormalizable theories is based on the power counting method, in the framework of the perturbation theory. Such issues have been
considered in the sixties and seventies (for a review see, for example, \cite{Pohlmeyer:1975gt}). As emphasized by several authors, what is lacking is the absence
of a natural prescription to make the solution unique.

\vspace{.1cm}

\noindent
A basic example of nonrenormalizable field theory is the one with exponential interaction for the scalar field $\phi$.
The potential
\be
\lambda \int_Id^Dx :\exp(\alpha \phi): \ ,
\ee
$\lambda\geq0$, with $I$ a cube in $\RR^D$, and with ultraviolet cutoff at distance
$\gamma^{-N}$, $\gamma>1$, $N>0$, has been investigated in several papers, in particular in \cite{Albeverio:1979nc}.
It turns out that for $D>2$, the Schwinger functions
converge to the free ones, for all $\alpha$ and for all $\lambda\geq0$. The same happens for $D=2$, but now
only in a given range of $\alpha$. Namely, there exists $\alpha_0>\sqrt{4\pi}$, such that for $\alpha>\alpha_0$ and for all $\lambda\geq0$, the Schwinger functions
converge to the free ones. For $D=2$ and $\alpha^2<4\pi$, the limit $I\to \RR^D$ gives nontrivial Schwinger functions.

\vspace{.1cm}

\noindent The essential point in the investigation of \cite{Albeverio:1979nc} is that
$\Delta_{\Lambda}(0)$, with $\Delta_{\Lambda}(x)$ the Feynman propagator
with cutoff on the momenta, grows sufficiently fast to kill the fluctuations of $\phi$, so that  $:\exp(\alpha \phi):= \exp(-{\alpha^2\over2}\Delta_{\Lambda}(0))\exp(\alpha \phi)$ vanishes in the limit
$\Lambda\to\infty$. We are not aware if such findings have been reproduced in the standard lattice regularization. It should be stressed that the
problem of removing the infrared cutoff in $D$ dimensional Euclidean space, that is of taking the infinite volume limit $I\to \RR^D$, has been addressed
by Raczka in \cite{Raczka:1978kq}. In particular, Raczka considered the exponential interaction with periodic boundary conditions and then considered the infinite volume limit.
It turns out that all the Wightman axioms are satisfied except the one concerning $SO(D)$ invariance that has not yet been proved. It should be stressed that without
periodic boundary conditions, translation invariance would be broken. A related work concerns
the spontaneous symmetry breaking (SSB) of space translations in Liouville theory, observed by D'Hoker, Freedman and Jackiw \cite{D'Hoker:1983ef}.
A key property of the SSB of space translations is that in this case one escapes a basic no-go theorem such as the one by Haag. Actually,
such a theorem essentially states that the interaction picture does not exist. It follows that perturbation theory,
where the interacting field is unitarily equivalent to the free field, is, in general, ill defined \cite{Haag,Haag:1992hx,Strocchi:2013awa}.
Another consequence of the SSB of space translations is that the K\"all\'en-Lehman spectral decomposition
\cite{Kallen:1952zz,Lehmann:1954xi} does not hold.

\vspace{.1cm}

\noindent
In this paper we investigate the scalar exponential interaction using the Schwinger's formalism.
The investigation, which is applied in the companion paper \cite{Matone:2015dya} concerning the formulation of the Higgs model in terms of exponential interaction,
is based on the observation that Schwinger's formalism allows us to use
$\exp(\alpha{\delta_{J(x)}})$
as the translation operator of $J(y)$ by the delta-source
\be
\alpha_x(y)\equiv \alpha\delta(y-x) \ .
\label{kjnoAbinitio}\ee
In particular, we will use the relation (see (\ref{iniziale}) for the notation)
 \be
 \exp\Big(\alpha{\delta\over \delta J(x)}\Big)\exp(-Z_0[J])=\exp(-Z_0[J+\alpha_x]) \ .
 \label{shift}\ee
The investigation can be generalized to linear combinations of exponential terms.

\vspace{.1cm}

\noindent
The paper is organized as follows.

\vspace{.1cm}

\noindent
In Sec. \ref{sec-exp}, after fixing the notation and shortly recalling the Schwinger formalism,
we show that the translation (\ref{shift}) is strictly related to the normal ordering of exponential operators. This naturally leads to
an alternative representation of the generating functional.
Furthermore, the term $\exp\big({\alpha^2\over 2}\Delta(0)\big)$, with $\Delta(x)$ the Feynman scalar propagator, is absorbed by the renormalization of $\mu$,
leading to scaling relations for the mass. In Eq.(\ref{further}) we introduce a modified Feynman propagator in the massless case,
whose dependence on the free parameter $m_0$ should be compared with the
ambiguities of taking the $D\to 4^+$ limit in dimensional regularization. In Eq.(\ref{gf}) we report scaling relations of the generating functionals $W[J]$.
We then discuss the SSB and its connection with the choice of the boundary conditions.

\vspace{.1cm}

\noindent
In Sec. \ref{sec-master} we consider the exponential interaction as a master potential to generate
other interactions. This is very natural in the case of polynomial interactions. In particular,
we provide a formulation of perturbation theory for the normal ordered potentials ${\lambda\over n!}:\phi^n:$
which is based on the potential $\mu^D:\exp(\alpha\phi):$.
It turns out that translations of $J(y)$ by $\alpha_x(y)$ generate the full terms related to normal ordering. This allows us
to absorb at once the $\Delta(0)$-terms which are typical of non-normal ordered potentials. Furthermore, in this approach, the functional derivatives with respect to $J$
to compute the generating functional are replaced by ordinary derivatives with respect to auxiliary parameters.
We then derive the explicit full perturbation series of the generating functional associated to the potentials ${\lambda\over n!}:\phi^n:$. Such an expression is reported in Eq.(\ref{centraltre}).
 Furthermore, as an illustration of
the method, we derive such a series even in the case of the one- and two-point functions.
We focus on scalar theories, but the method is general and similar investigations extend to other theories.

\vspace{.1cm}

\noindent
Section \ref{sec-conc} is devoted to the conclusions and to some suggestions for further developments of the proposed formulation of quantum field perturbation theory,
based on the exponential interaction, considered as a master potential.

\section{Exponential interaction}\label{sec-exp}

In this section we first introduce some notation and recall few facts on the Schwinger formalism. Next, we
use the translation operator $\exp(\alpha{\delta\over \delta J(x)})$ to express the generating functional $W[J]$
associated to the exponential interaction $\mu^D e^{\alpha\phi}$ as a power series in $\mu^D$. Such an analysis will shed light on some
properties of the exponential interaction. In particular, it turns out that the action
of $\exp(\alpha{\delta\over \delta J(x)})$, equivalent to introduce infinitely many charges, is strictly related to
normal ordering. This naturally leads to introduce an alternative expression for $W[J]$.
Such a little-known expression extends to generating functionals associated to arbitrary potentials \cite{FriedBook} and leads to some new relations \cite{Matone:2015aib}.
Next, we derive scaling relations for the mass, both in the dimensional and in
the momentum cutoff regularization.
Then, we derive a scaling law relating $W[J]$ at different values of $\mu$. We conclude this section by discussing the
possible SSB of space-time translations of the exponential interaction, relating it to the choice of the boundary conditions.

\subsection{Notation}\label{sec-thetwo}

In the following we set the notation, essentially the one in Ramond's book \cite{Ramond:1981pw}, and recall few facts on the Schwinger's formalism.

\noindent
In $D$ dimensional Euclidean space, the generating functional is defined by
\be
W[J]=e^{-Z[J]}=N\int D\phi \exp\Big[-\int d^Dx \Big({1\over2}\partial_\mu\phi\partial_\mu\phi + {1\over2}m^2\phi^2 +V(\phi)-J\phi\Big)\Big] \ .
\label{primordiale}\ee
$Z[J]$ generates the connected $N$-point functions
\be
(-1)^N{\delta^N Z[J]\over \delta J(x_1)\ldots \delta J(x_N)}|_{J=0}=\langle 0|T\phi(x_1)\ldots\phi(x_N)|0\rangle_c \ .
\label{initial}\ee
Shifting $\phi$ in the term $J\phi$ in (\ref{primordiale}) by an arbitrary function $f$, gives, on the right-hand side of (\ref{initial}), the correlators of $\phi+f$.
Such a shift is equivalent to replace $Z[J]$ by $Z[J]-\int d^Dx J(x)f(x)$,
so that, for $N\geq 2$,
\be
{\delta^N (Z[J]-\int d^Dx J(x)f(x))\over \delta J(x_1)\ldots \delta J(x_N)}={\delta^N Z[J]\over \delta J(x_1)\ldots \delta J(x_N)} \ .
\label{duetre}\ee
This implies that the $N$-point functions of $\phi$ and $\phi+f$ coincide for $N\geq2$.
In particular, choosing $f(x)=-\langle\phi(x)\rangle$, one sees
that the connected $N$-point functions of $\phi$ and $\eta=\phi-\langle\phi\rangle$ coincide for $N\geq2$
\be
\langle 0|T\phi(x_1)\ldots\phi(x_N)|0\rangle_c=\langle 0|T\eta(x_1)\ldots\eta(x_N)|0\rangle_c \ ,
\label{duedue}\ee
which holds even when $\langle\phi(x)\rangle$ depends on $x$.
A simple example to see that factorized terms disappear from the connected $\phi$ correlators, even when $\langle\phi(x)\rangle$ is not vanishing, is provided by
the expression  of $\langle0|T\phi(x_1)\phi(x_2)|0\rangle_c$ in the case $V=\lambda\phi$. By (\ref{duepuntinotevole}), one immediately sees that
the term $-\langle\phi(x_1)\rangle\langle\phi(x_2)\rangle=-\lambda^2/m^4$ is canceled by the other $\lambda^2$-term.

\noindent
Set
\be
\langle f(x_1,\ldots,x_n)\rangle_{x_j\ldots x_k}\equiv\int d^D x_j\ldots\int d^D x_k
f(x_1,\ldots,x_n) \ ,
\ee
and denote by $\langle f(x_1,\ldots,x_n)\rangle$ integration of $f$ over all the variables.
Schwinger's formalism is based on the relation
\be
W[J]=N e^{-\langle V({\delta\over\delta J})\rangle}e^{-Z_0[J]} \ ,
\label{diventa}\ee
where
\be
Z_0[J]=-{1\over2}\langle J(x)\Delta(x-y) J(y)\rangle \ ,
\ee
and
\be
\Delta(x-y)=\int {d^Dp \over (2\pi)^D} {e^{ip(x-y)}\over p^2+m^2} \ ,
\ee
is the Feynman propagator.
On the other hand, $Z[J]$ can be expressed in the form
\be
Z[J]=-\ln N+Z_0[J]-\ln(1+\delta[J]) \ ,
\ee
where
\be
\delta[J]=e^{Z_0[J]}\Big( e^{-\langle V({\delta\over\delta J})\rangle}-1\Big)e^{-Z_0[J]} \ .
\ee
Expanding $\delta[J]$ in the power series of the dimensionless coupling constant $\lambda$
\be
\delta[J]=\sum_{k=1}^\infty \delta_k[J]\lambda^k \ ,
\ee
leads to the perturbative series
\be
Z[J]=-\ln N+Z_0[J]-\lambda \delta_1[J]-\lambda^2\Big(\delta_2[J]-{1\over2}\delta_1^2[J]\Big)+\ldots \ .
\ee

\subsection{Generating functional from the translation operator}

Consider the potential
\be
V(\phi)=\mu^D e^{\alpha\phi} \ ,
\ee
where $\mu$ and $\alpha$ have mass dimension 1 and $(2-D)/2$, respectively.
Dropping the constant $N$ in the expression of $W[J]$, we have
\begin{align}
W[J] & = \exp\Big[-\mu^D\langle \exp (\alpha {\delta\over \delta J})\rangle \Big] \exp(-Z_0[J])\cr
&=\sum_{k=0}^\infty {(-\mu^{D})^k\over k!} \langle \exp(\alpha {\delta\over\delta J})\rangle^k \exp(-Z_0[J]) \ .
\label{Winiziale}\end{align}
Since $\exp(\alpha{\delta\over \delta J(x)})$ translates $J(y)$ by
\be
\alpha_x(y)\equiv \alpha\delta(y-x) \ ,
\label{pointlikesource}\ee
it follows that in this case
Eq.(\ref{diventa}) can be easily implemented.
In particular,
\be
\exp(\alpha{\delta\over \delta J(x)})\exp(-Z_0[J])=\exp(-Z_0[J+\alpha_x])\exp(\alpha{\delta\over \delta J(x)})=\exp(-Z_0[J+\alpha_x]) \ ,
\ee
where
\begin{align}
Z_0[J+\alpha_x]&= -{1\over 2}\int d^Dy\int d^D z (J(y)+\alpha\delta(x-y))\Delta(y-z)(J(z)+\alpha\delta(x-z))\cr
&=Z_0[J]-{\alpha^2\over2}\Delta(0)-\alpha \int d^Dy J(y)\Delta(y-x) \ .
\label{iniziale}\end{align}
Therefore,
\be
W[J]=\sum_{k=0}^\infty {(-\mu^{D})^k\over k!}\langle\exp(-Z_0[J+\alpha_{x_1}+\ldots+\alpha_{x_k}])\rangle \ ,
\label{espansione}\ee
or, more explicitly,
\begin{align}
&W[J]=\exp(-Z_0[J])\sum_{k=0}^\infty{(-\mu^{D})^k\over k!}\exp\Big({k\alpha^2\over2}\Delta(0)\Big)G_k[J] \ ,
\label{equation}\end{align}
with $G_0=1$ and
\be
G_k[J]=\int d^Dz_1\ldots\int d^D z_k\exp\Big( \alpha\int d^Dz J(z)\sum_{j=1}^k \Delta(z-z_j)+\alpha^2\sum_{l>j}^k\Delta(z_j-z_l)\Big) \ ,
\ee
$k\in\NN_+$, where the second summation starts contributing from $k=2$.

\noindent
Since the derivatives of $W$ with respect to $J$ commute to the right of the
first exponential in (\ref{Winiziale}), it follows that even the $N$-point functions $G^{(N)}(x_1,\ldots,x_N)$ are obtained by acting with the translation operator. In this case, such operators act on
\be
F[J,\alpha,x_1,\ldots,x_N]=(-1)^N{\delta^N \exp(-Z_0[J])\over \delta J(x_1) \ldots \delta J(x_N)} \ ,
\ee
so that
\be
G^{(N)}(x_1,\ldots,x_N)=\sum_{k=0}^\infty {(-\mu^{D})^k\over k!}\langle F[\alpha_{z_1}+\ldots+\alpha_{z_k},\alpha,x_1,\ldots,x_N]\rangle_{z_1\ldots z_k} \ ,
\ee
where $\alpha_z$ is the delta source defined in (\ref{pointlikesource}).

\subsection{An alternative representation of $W[J]$}

Here we show that the action of the translation operator is strictly related to the normal ordering. We will see that such an analysis
will naturally lead to introduce an alternative representation of $W[J]$.

\noindent
Let us first observe that the generating functional can be expressed as the vacuum expectation value (vev)
\be
W[J]=\langle 0| T \exp\Big[\int d^Dx\Big(-\mu^De^{\alpha\phi(x)}+J(x)\phi(x)\Big)\Big]|0\rangle \  ,
\label{tredodici}\ee
where here $|0\rangle$ is the vacuum of the free theory.
Setting
\be
\phi(z)=\phi\,'(z)+\int d^Dy J(y)\Delta(y-z)
\ee
in (\ref{tredodici}) and then removing the prime in $\phi\,'$, yields
\be
W[J]=e^{-Z_0[J]}\langle 0 | T \exp\Big[-\mu^D\int d^Dz e^{\alpha(\phi(z)+\int d^Dy J(y)\Delta(y-z))}\Big]|0\rangle \ .
\label{showsthat}\ee
We now show that the terms $e^{{k\alpha^2\over2}\Delta(0)}$,  $e^{\alpha^2\sum_{l>j}^k\Delta(z_j-z_l)}$ and
$e^{\alpha\langle J(z)\sum_{j=1}^k \Delta(z-z_j)\rangle_{z}}$ in (\ref{equation}), generated by the shift $J(y)\to J(y)+\alpha\delta(y-x)$,
are the same as the ones due to the normal ordering of the exponentiated operators.
To check this, note that (\ref{Winiziale}) corresponds to the power expansion of
$e^{-\langle V(\phi)\rangle}$ in (\ref{tredodici}), that is,
\be
W[J]= \sum_{k=0}^\infty {(- \mu^{D})^k\over k!} \int d^Dz_1\ldots \int d^D z_k \langle0| T e^{\alpha\phi(z_1)}\ldots e^{\alpha\phi(z_k)}e^{\int d^D z J(z)\phi(z)}|0\rangle \ .
\label{riottieni}\ee
Next, recall that
\be
:\exp(\alpha \phi(z)):=\exp\Big(-{\alpha^2\over 2}\Delta(0)\Big)\exp(\alpha \phi(z)) \ ,
\label{exponentiated}\ee
and
\be
T\exp(\int d^Dz J(z)\phi(z))=\exp(-Z_0[J]):\exp(\int d^Dz J(z)\phi(z)): \ .
\label{Tdi}\ee
Furthermore,
\begin{align}
T&:\exp(\alpha \phi(z_1)): \ldots :\exp(\alpha \phi(z_k))::\exp(\int d^Dz J(z)\phi(z)):\cr
=&\exp\big(\alpha^2 \sum_{l>j}^k\Delta(z_j-z_k)+\alpha\int d^Dz J(z)\sum_{j=1}^k \Delta(z-z_j)\big) \cr
&:\exp(\alpha \phi(z_1))\ldots \exp(\alpha \phi(z_k))\exp(\int d^Dz J(z)\phi(z)): \ .
\label{eck}\end{align}
Equations (\ref{exponentiated}), (\ref{Tdi}), and (\ref{eck}) imply
\begin{align}
 T & \exp({\alpha  \phi(z_1)})\ldots \exp({\alpha\phi(z_k)})\exp({\int d^D z J(z)\phi(z)}) \cr
   = & T\big[\exp({\alpha\phi(z_1)})\ldots \exp({\alpha\phi(z_k)}) T \exp({\int d^D z J(z)\phi(z)})\big] \cr
= & \exp(-Z_0[J])\exp[{{\alpha^2}({k\over 2}\Delta(0)+\sum_{l>j}^k \Delta(z_j-z_l))+ \alpha\int d^Dz J(z)\sum_{j=1}^k \Delta(z-z_j)}]\cr
& :\exp(\alpha\phi(z_1))\ldots \exp({\alpha\phi(z_k)})\exp({\int d^D z J(z)\phi(z)}) :  \ .
 \label{riottienidue}\end{align}
Since the vev of the last normal ordering is one, it follows that (\ref{riottieni}) reproduces, by (\ref{riottienidue}), Eq.(\ref{equation}).
We note that a slightly modified version of (\ref{riottienidue}) provides an alternative way to prove Eq.(\ref{showsthat}). First note that the vev of (\ref{riottienidue}) coincides with the one of
\begin{align}
  \exp(-Z_0[J]) & \exp\Big(\alpha\int d^Dz J(z)\sum_{j=1}^k \Delta(z-z_j)\Big) T  \exp({\alpha  \phi(z_1)})\ldots \exp({\alpha\phi(z_k)}) \cr
= & \exp(-Z_0[J])\exp[{{\alpha^2}({k\over 2}\Delta(0)+\sum_{l>j}^k \Delta(z_j-z_l))+ \alpha\int d^Dz J(z)\sum_{j=1}^k \Delta(z-z_j)}]\cr
& :\exp(\alpha\phi(z_1))\ldots \exp({\alpha\phi(z_k)}):  \ .
 \label{riottienitree}\end{align}
Equation (\ref{showsthat}) then follows by replacing the operator in the vev of (\ref{riottieni}) by (\ref{riottienitree}).

\noindent Let us show that $W[J]$ admits a representation which is an alternative to the representation (\ref{Winiziale}).
The key point is the following relation
\begin{align}
\exp  & \Big[{1\over2}\int d^D x_1\int d^D x_2 {\delta\over \delta J(x_1)} \Delta^{-1}(x_1-x_2){\delta\over\delta J(x_2)}\Big]\exp\Big(\alpha\int d^Dz J(z)\sum_{j=1}^k \Delta(z-z_j)\Big) \cr
&=\exp[{{\alpha^2}({k\over 2}\Delta(0)+\sum_{l>j}^k \Delta(z_j-z_l))+ \alpha\int d^Dz J(z)\sum_{j=1}^k \Delta(z-z_j)}] \ ,
\label{key}\end{align}
where
$\Delta^{-1}(y-x)$ is the inverse of the Feynman propagator
\be
\int d^D z \Delta^{-1}(x-z)\Delta(z-y)=\delta(x-y) \ ,
\ee
that is
\be
\Delta^{-1}(y-x)=\int {d^Dp\over (2\pi)^D}(p^2+m^2) e^{ip(y-x)} \ .
\ee
By (\ref{equation}) it then follows
\begin{align}
W[J]=&\exp(-Z_0[J])\exp\Big({1\over2}\int d^D x_1\int d^D x_2 {\delta\over \delta J(x_1)} \Delta^{-1}(x_1-x_2){\delta\over\delta J(x_2)}\Big) \cr
&\exp\Big[ -\mu^D \int d^Dx \exp\Big(\alpha\int d^Dz J(z)\Delta(z-x)\Big)\Big] \ .
\end{align}
It turns out that such a little-known expression extends to generating functionals associated to arbitrary potentials \cite{FriedBook}. Recently, in \cite{Matone:2015aib}, it has been shown that it leads to
some new relations.

\subsection{Mass renormalization and scaling}

\noindent The term $\exp\big({k\alpha^2\over2}\Delta(0)\big)$ in (\ref{equation}) can be absorbed either by considering the normal ordered potential $\mu^D:\exp(\alpha\phi):$ or by redefining $\mu$
\be
\mu^D=\mu_0^D \exp\Big(-{\alpha^2\over 2}\Delta(0)\Big) \ .
\label{duetresei}\ee
Since in dimensional regularization one has \cite{'tHooft:1972fi}
\be
\Delta(0)={m^{D-2}\over (4\pi)^{D/2}}\Gamma(1-D/2) \ ,
\label{mamanddad}\ee
it follows that (\ref{duetresei}) is equivalent to the scaling relation
\be
m^{D-2}={2D(4\pi)^{D/2} \over \alpha^2 \Gamma(1-D/2)}\ln{\mu_0\over \mu} \ .
\label{questanelleconclusioni}\ee
Recall that $\Delta(0)$ is singular for $D=2n$, $n=\NN_+$.
In particular, its expansion near to $D=4$ reads
\be
\int {d^Dp \over (2\pi)^D} {1\over p^2+m^2}={m^2\over (4\pi)^2}\Big({2\over D-4}-\psi(2)\Big)+{\cal O}(D-4) \ ,
\label{van}\ee
where
$\psi(2)={3\over2}-\gamma$,
with $\gamma=0.5772...$, the Euler-Mascheroni constant.
Dropping the terms in (\ref{van}) vanishing for $D=4$, yields
\be
m^2={4(4\pi)^2(D-4)\over \alpha^2[2-(D-4)\psi(2)]}\ln {\mu_0^2\over\mu^2} \ .
\ee
We note that setting
\be
m^2=(D-4)m_0^2 \ ,
\label{sceltainusuale}\ee
gives, in the $D\to4$ limit, the finite mass
\be
m_0^2={32\pi^2\over \alpha^2}\ln {\mu_0^2\over\mu^2} \ .
\ee
{}From the point of view of the Feynman propagator
\be
\Delta_{m_0}(x-y):=\int {d^D p\over(2\pi)^D}{e^{ip(x-y)}\over p^2+(D-4)m_0^2} \ ,
\label{further}\ee
it should correspond, for $D=4$, to a massless particle. In this respect it is interesting to compare
\be
\Delta_{m_0}(0)={m_0^2\over 8\pi^2}
\label{furthersss}\ee
with \cite{'tHooft:1972fi}
\be
\int {d^Dp\over (2\pi)^D}(p^2)^k=0 \ ,
\label{critical}\ee
holding for $k,D\in\CC$, except in the cases $D=2n$, $n=\NN_+$, where there are some open questions \cite{Leibbrandt:1975dj}.
On the other hand, adding a term $(D-4)m_0^2\phi^2/2$ in the path-integral formulation is in the spirit of dimensional
regularization. The free parameter $m_0$ would be an alternative to the ambiguities
 stressed in \cite{Leibbrandt:1975dj}, where it is shown that, depending on the calculation method, the $D\to 4^+$ limit yields either infinity, zero or a finite value.

\noindent The above analysis can be compared with the regularization by the momentum cutoff $\Lambda$. For $D=4$ we have
\be
\int_\Lambda{d^4p\over(2\pi)^4}{1\over p^2+m^2}={m^2\over 16\pi^2}\Big[{\Lambda^2\over m^2}-\ln{\Lambda^2\over m^2}\Big]+{\cal O}\big[(\Lambda^{-1})^0\big] \ ,
\ee
which implies the scaling relation
\be
\mu^4=\mu_0^4 \Big({\Lambda^2\over m^2}\Big)^{\alpha^2m^2\over  32\pi^2}\exp\Big({-{\alpha^2\Lambda^2\over 32 \pi^2 }}\Big) \ .
\ee
Note that setting
\be
\alpha=\pm {8\pi\over m} \ ,
\ee
yields a simple scaling relation involving only the dimensionless ratios $\Lambda/m$ and $\mu/\mu_0$
\be
{\mu\over\mu_0}={\Lambda\over m}\exp\Big[-{1\over2}\Big({\Lambda\over m}\Big)^2\Big] \ .
\ee
We note that rescaling by the normal ordering contributions has been considered in several contexts, for example in the framework of Liouville theory \cite{D'Hoker:1983ef}. This is related to
the scaling law for $W[J]$, investigated in the next subsection.

\subsection{The scaling of $W[J]$ and $\langle\phi(x)\rangle$}

Note that by (\ref{Winiziale}) or, equivalently, by (\ref{tredodici}), it follows that $W[J]$ satisfies the equations
\be
\Big({\partial\over\partial \mu_0^D}+\int d^D x \exp\Big(\alpha{\delta\over\delta J(x)}\Big)\Big)W[J]=0 \ ,
\label{by}\ee
and
\be
\Big({\partial\over\partial \alpha}+\mu_0^D\int d^D x \exp\Big(\alpha{\delta\over\delta J(x)}\Big){\delta\over\delta J(x)}\Big)W[J]=0 \ .
\label{bye}\ee
Another equation satisfied by $W[J]$ follows by noticing that since
\be
\langle \Delta(x-y)\rangle_x={1\over m^2} \ ,
\label{usingagain}\ee
one has
\be
\int d^Dx{\delta G_k[J]\over\delta J(x)}={k\over m^2} G_k[J] \ .
\ee
By (\ref{equation}) such a relation implies
\be
{\alpha\over m^2} {\partial\ln W[J]\over\partial \ln \mu_0^D}=\int d^Dx \Big({\delta\ln W[J]\over \delta J(x)}-{J(x)\over m^2}\Big) \ .
\label{rilevantibus}\ee
This equation can be also obtained by (\ref{by}),
together with the integrated version of the Schwinger-Dyson equation
\be
\Big(\int d^Dy\Delta^{-1}(y-x){\delta\over\delta J(y)}+\int d^Dy{\delta V\over \delta\phi(x)}\Big({\delta\over\delta J(y)}\Big)-J(x)\Big)W[J]=0 \ .
\label{lastandard}\ee
Note that Eq.(\ref{rilevantibus}) is equivalent to the scaling relation
\be
W[J,{\mu_0^D}]=W[J,{\bar\mu_0^D}]\exp\Big[{m^2\over \alpha}\int^{\mu_0^D}_{\bar\mu_0^D}{du\over u}\int d^D x \Big(\phi_{{\rm cl}\, u}(x)-{J(x)\over m^2}\Big)\Big] \ ,
\label{gf}\ee
where ${\mu_0^D}$ and ${\bar\mu_0^D}$ are two arbitrary values of the mass scale and
\be
\phi_{{\rm cl}\, u}(x)={\delta\ln W[J,u]\over \delta J(x)} \ .
\ee
Since $W[J,0]$ is the free generating functional, Eq.(\ref{gf}) implies the following representation for $W[J,{\mu_0^D}]$
\be
W[J,{\mu_0^D}]=\exp(-Z_0[J])\exp\Big[{1\over \alpha}\int^{\mu_0^D}_{0}{du\over u}\int d^D x \Big(m^2\phi_{{\rm cl}\, u}(x)-{J(x)}\Big)\Big] \ .
\label{gfnata}\ee
Taking the functional derivatives of Eq.(\ref{gf}) with respect to $J$, evaluated at $J=0$, generates identities between the Green functions at different
scales. For example,
\be
\langle\phi(x)\rangle_{\mu_0^D}-\langle\phi(x)\rangle_{\bar\mu_0^D}={m^2\over\alpha}\int^{\mu_0^D}_{\bar\mu_0^D}{du\over u}\int d^D y\langle 0 | T\phi(x)\phi(y)|0\rangle_{u}+{D\over\alpha}\ln{\bar\mu_0\over\mu_0} \ ,
\ee
where
\be
\langle\phi(x)\rangle_{u}= {\delta \ln W[J,{u}]\over \delta J(x)}|_{J=0} \ ,
\ee
and
\be
\langle 0 | T\phi(x)\phi(y)|0\rangle_{u}={\delta^2 \ln W[J,{u}]\over \delta J(x)\delta J(y)}|_{J=0} \ .
\ee

\noindent
Let us now consider the one-point function
\be
\langle \phi(x)\rangle={\alpha\sum_{k=1}^\infty\Big[ {(-\mu_0^D)^k\over k!}
\langle\exp \Big(\alpha^2\sum_{l>j}^k\Delta(z_j-z_l)\Big)\sum_{j=1}^k \Delta(x-z_j)\rangle_{z_1\ldots z_k}\Big]\over
\sum_{k=0}^\infty{(-\mu_0^D)^k\over k!}
\langle\exp \Big(\alpha^2\sum_{l>j}^k\Delta(z_j-z_l)\Big)\rangle} \ .
\label{wonderfull}\ee
Such an expression is formally translation invariant; therefore it should be treated as a constant. However, it needs a finite volume regularization.
The infrared regularization for the exponential interaction has been investigated, e.g., in \cite{Albeverio:1979nc} and by
Raczka in \cite{Raczka:1978kq}, who considered periodic boundary conditions that may break the $SO(D)$ invariance. A related issue
concerns the SSB of space translations, in the case of Liouville theory, observed by D'Hoker, Freedman,
and Jackiw in \cite{D'Hoker:1983ef}.

\noindent
We conclude this section by observing that the above analysis can be extended to the case of more general potentials, such as
\be
V(\phi)=\sum_{k=1}^N\mu_k^D \exp({\alpha_k\phi}) \ ,
\label{potentials}\ee
whose corresponding generating functionals are
\be
W[J]=\sum_{k=0}^\infty{(-1)^k\over k!}\langle \sum_{j=1}^N\mu_j^D\exp\big(\alpha_j{\delta\over\delta J}\big)\rangle^k\exp(-Z_0[J]) \ .
\ee
We also note that interesting cases concern the extension to more
scalar fields with exponential interactions.

\section{Exponential interaction as master potential}\label{sec-master}

In this section we consider
the exponential interaction as a master potential to get the generating functional for polynomial interactions.
After introducing the method, we will derive the explicit expression of the functional generator in the case of the normal ordered potentials ${\lambda\over n!}:\phi^n:$.
This is obtained by deriving $\mu^D:\exp(\alpha\phi):$ with respect to $\alpha$, so that all the $\Delta(0)$-terms due to the non normal ordered contributions are absorbed at once from the very beginning.
We note that whereas in the standard approach
the generating functional is obtained by taking the functional derivatives with respect to $J$, here such derivatives are replaced by standard derivatives
with respect to auxiliary parameters.
We conclude with illustrative examples by computing the one- and two-point functions.

\subsection{The method}

The starting point is to note that
\be
\phi^n= \partial_\alpha^n e^{\alpha\phi}|_{\alpha=0} \ .
\ee
It follows that the generating functional corresponding to $V={\lambda\over n!}\phi^n$ is the following modified version
of (\ref{Winiziale})
\begin{align}
W^{(n)} [J] & =  \exp\Big[-{\lambda\over n!}\partial_\alpha^n \langle \exp \alpha {\delta\over\delta J}\rangle \Big] \exp(-Z_0[J])|_{\alpha=0}\cr
&=\sum_{k=0}^\infty {(-\lambda)^k\over k!n!^k}\partial_{\alpha_1}^n\ldots\partial_{\alpha_k}^n\langle \exp \alpha_1 {\delta\over\delta J}\rangle\ldots \langle \exp \alpha_k {\delta\over\delta J}\rangle \exp(-Z_0[J])|_{\alpha^{(k)}=0} \ ,
\label{Wphin}\end{align}
where $\alpha^{(k)}:=(\alpha_1,\ldots,\alpha_k)$.
We now show that, within such an approach, the $\Delta(0)$-terms can be absorbed at once. First note that by (\ref{exponentiated})
\be
:\phi(x)^n:=\partial_\alpha^n :e^{\alpha\phi(x)}:|_{\alpha=0}=\sum_{k=0}^n\Big(\begin{array}{c}n\\ k
\end{array}\Big) \partial_\alpha^k e^{-{\alpha^2\over2}\Delta(0)}|_{\alpha=0}\phi(x)^{n-k} \ .
\label{ilNO}\ee
The first few cases are,
\begin{align}
:\phi^2(x):&=\phi^2(x)-
\Delta(0) \ , \cr
:\phi^3(x):&=\phi^3(x)-3\Delta(0)\phi(x) \ , \cr
:\phi^4(x):&=\phi^4(x)-6\Delta(0)\phi^2(x)+3\Delta^2(0) \ .
\label{seix}\end{align}
It follows by (\ref{ilNO}) that the generating functional associated to $V={\lambda\over n!}:\phi^n:$
corresponds to (\ref{Wphin}), where now each $\langle\exp \alpha_k {\delta\over\delta J}\rangle$
is multiplied by $\exp\big(-{\alpha_k^2\over2}\Delta(0)\big)$. The net effect on $W^{(n)} [J]$ in (\ref{Wphin}) is that now, acting with the translation operator
$\langle\alpha{\delta\over \delta J}\rangle$
on
$Z_0[J]$, one may replace $Z_0[J+\alpha_x]$ in (\ref{iniziale}) by
\be
\tilde Z_0[J+\alpha_x]=Z_0[J]-\alpha \int d^Dy J(y)\Delta(y-x) \ .
\label{mediana}\ee
This leads to the central result for the generating functionals
\begin{align}
& W^{(n)}[J] = \exp(-Z_0[J])\sum_{k=0}^\infty{(-\lambda)^k\over (n!)^k k!}\partial_{\alpha_1}^n\ldots\partial_{\alpha_k}^n \cr
&\int d^Dz_1\ldots\int d^D z_k \exp\Big(\int d^DyJ(y)\sum_{j=1}^k\alpha_j\Delta(y-z_j)+\sum_{l>j}^k\alpha_j\alpha_l\Delta(z_j-z_l)\Big)|_{\alpha^{(k)}=0}\ . \cr
\label{central}\end{align}
There is a nice property satisfied by the $N$-point functions. Namely, since the derivatives with respect to $J$ of $W^{(n)} [J]$
commute with the derivatives with respect to the $\alpha_j$'s, and since the $N$-point functions are obtained at $J=0$, it follows that the term
$\langle J(y)\sum_{j=1}^k\alpha_j\Delta(y-z_j)\rangle_y$ in (\ref{central}) can be set to zero before computing the derivatives with respect to the $\alpha_j$'s. Also note that
the functional derivatives to get the correlators with respect to $J$ are easily computed. In particular, the only contributions
come from the free part, i.e., $Z_0[J]$, and from the term $\langle J(y)\sum_{j=1}^k\alpha_j\Delta(y-z_j)\rangle_y$, that contributes
by a term $\sum_{j=1}^k\alpha_j\langle\Delta(x-z_j)\rangle_y$ for each derivative with respect to $J(x)$.

\subsection{$W^{(n)}[J]$}\label{subsec-W}

Here we derive the explicit expression of $W^{(n)}[J]$. It is instructive to start by considering the simpler case of $W^{(n)}[J]$ at $J=0$
\be
 W^{(n)}[0] = \sum_{k=0,k\neq1}^\infty{(-\lambda)^k\over n!^k k!}\partial_{\alpha_1}^n\ldots\partial_{\alpha_k}^n
\langle e^{\sum_{l>j}^k\alpha_j\alpha_l\Delta(z_j-z_l)}\rangle|_{\alpha^{(k)}=0} \ .
\label{central2}\ee
When $kn$ is odd, there are no contributions to the $k$th term of the series. When $kn$ is even, $\partial_{\alpha_1}^n\ldots\partial_{\alpha_k}^n$ selects,
by setting $\alpha^{(k)}=0$, $n!^k$ times the coefficient of $\alpha_1^n\cdots\alpha_k^n$ in the expansion of the exponential. In order to investigate
the contributions to $W^{(n)}[0]$ coming from such an expansion, it is useful to
consider the multinomial identity
\be
{1\over ({kn\over 2})!}\Big(\sum_{l>j}^k\alpha_j\alpha_l\Delta(z_j-z_l)\Big)^{kn\over2}=\sum_{\sum_{l>j}^k m_{jl}=kn/2} {1\over \prod_{l>j}^km_{jl}!}\prod_{l>j}^k(\alpha_j\alpha_l\Delta(z_j-z_l))^{m_{jl}} \ ,
\label{seidieci}\ee
where $0\leq m_{jl}\leq kn/2$.
For each $l=1,\ldots,k$, the total exponent of $\alpha_l$ in (\ref{seidieci})  is
\be
p_l:=\sum_{i=1}^{l-1}m_{il}+\sum_{j=l+1}^k m_{lj} \ ,
\label{pelle}\ee
where it is understood that for $l=1$ there is only the second summation, while for $l=k$ there is only the first one. The contributions to
the $k$th term in the series (\ref{central2}) are the ones with
\be
p_1=\ldots=p_k=n \ ,
 \label{pio}\ee
which gives $\sum_{l=1}^kp_l=kn$. On the other hand,  Eq.(\ref{pelle}) implies  $\sum_{l=1}^kp_l=2\sum_{l>j}^km_{jl}$, so that the condition (\ref{pio}) includes the condition $\sum_{l>j}^k m_{jl}=kn/2$
 reported in the summation in the right-hand side of (\ref{seidieci}).
Hence,
\begin{align}
W^{(n)}[0] = \sum_{k=0,k\neq1}^\infty{{(-\lambda)^k\over k!}}\sum_{p_1=\ldots =p_k=n} [k| m] \prod_{l>j}^k\langle\Delta(z_j-z_l)^{m_{jl}}\rangle\ ,
\label{centraldue}\end{align}
where for $kn$ even
\be
[k|m]:= {1\over \prod_{l>j}^km_{jl}!}\ ,
\label{edh}\ee
and $[k|m]=0$ otherwise. It is everywhere understood in the paper that, for $k=0,1$, terms of the form $\prod_{l>j}^ka_{jl}$
are set to 1.

\noindent
Let us extend the analysis to the case of $W^{(n)}[J]$. First, for each $k$ in the series (\ref{central}),
consider the following identity satisfied by the term in the expansion of the exponential, in the integrand of
$W^{(n)}[J]$, containing $\alpha_1^n\cdots\alpha_k^n$,
unless in the case $J=0$ with $kn$ odd,
\begin{align}
\sum_{p=0}^{[{kn\over2}]} &  {1\over (kn-2p)!}{1\over p!} \langle J(y)\sum_{i=1}^k\alpha_i \Delta(y-z_i)\rangle_y^{kn-2p}\Big(\sum_{l>j}^k\alpha_j\alpha_l \Delta(z_j-z_l)\Big)^p \cr
= & \sum_{p=0}^{[{kn\over2}]}  \sum_{\sum_{i=1}^k q_i=kn-2p}\sum_{\sum_{l>j}^k m_{jl}=p}{1\over \prod_{i=1}^kq_i! \prod_{l>j}^km_{jl}!} \cr
&\prod_{i=1}^k\langle J(y)\alpha_i\Delta(y-z_i)\rangle_y^{q_i}\prod_{l>j}^k\Big(\alpha_j\alpha_k\Delta(z_j-z_l)\Big)^{m_{jl}} \ ,
\label{power}\end{align}
where $[a]$ denotes the integer part of $a$, $0\leq q_i\leq kn-2p$, $0\leq m_{jl}\leq p$. Let us first show how this formula reproduces, for $J=0$, the identity (\ref{seidieci}).
If $kn$ is even, then, for $J=0$, all the terms $\langle J(y)\alpha_i\Delta(y-z_i)\rangle_y^{q_i}$, $i=0,\ldots,k$, in the right-hand side of (\ref{power}), are zero unless $q_i=0$. Similarly,
for $J=0$ the term $\langle J(y)\sum_{i=1}^k\alpha_i \Delta(y-z_i)\rangle_y^{kn-2p}$, on the left-hand side of (\ref{power}), contributes only for $kn-2p=0$; this is, of course, consistent
with the condition $\sum_{i=1}^k q_i=kn-2p$.
Therefore, for $J=0$ the summation over $p$ in (\ref{power}) reduces to the term with $p={kn\over2}$. Since $q_1=\ldots=q_k=0$, we have $\prod_{i=1}^kq_i!=1$, and (\ref{power}) reduces to (\ref{seidieci}).
If $kn$ is odd and $J=0$, then the $k$th term in (\ref{central}) is zero. This follows also by  (\ref{power}). In this case, the only possible contribution to  (\ref{power})
 would come from $q_1=\ldots=q_k=0$. On the other hand,
this would give $\sum_{i=1}^k q_i=0$ that cannot be equal to the odd number $kn-2p$.
Therefore, such a configuration of the $q_i$s is not included in the summation.

\noindent
Let us now consider (\ref{power}) for arbitrary $J$. For each $l=1,\ldots,k$, and for each choice of $m_{jl}$'s, the total exponent of $\alpha_l$ in (\ref{power}) is
\be
p_l:=\sum_{i=1}^{l-1}m_{il}+\sum_{j=l+1}^k m_{lj}+q_l \ .
\label{pelleforj}\ee
The two conditions in the summation's indices, in the right-hand side of (\ref{power}), imply
\be
\sum_{l=1}^kp_l=2\sum_{l>j}^k m_{jl}+\sum_{l=1}^k q_l=2p+kn-2p=kn  \ .
\label{soddisfatto}\ee
For each $k$, the only contributions to $W^{(n)}[J]$ are the ones with $p_1=\ldots=p_k=n$. This condition implies (\ref{soddisfatto}), so that
\begin{align}
W^{(n)}[J] = & e^{-Z_0[J]}\sum_{k=0}^\infty{{(-\lambda)^k\over k!}}\sum_{p=0}^{[{kn\over2}]} \sum_{\sum_{i=1}^k q_i=kn-2p}\sum_{p_1=\ldots=p_k=n}
 \cr
& [k|m,q]\prod_{i=1}^k\langle\langle J(y)\Delta(y-z_i)\rangle_y^{q_i}\prod_{l>j}^k\Delta(z_j-z_l)^{m_{jl}}\rangle\ .
\label{centraltre}\end{align}

\subsection{One- and two-point functions}

\noindent As an illustration of the method, we now derive the one- and two-point functions. This can be done directly considering the functional derivatives of (\ref{centraltre}); nevertheless it is
instructive to investigate the combinatorics starting again from the expression (\ref{central}).

\noindent Let us start with the one-point function
\begin{align}
\langle\phi(x)\rangle &={1\over  W^{(n)} [0]}\sum_{k=1}^\infty{(-\lambda)^k\over k!}\partial_{\alpha_1}^n\ldots\partial_{\alpha_k}^n \cr
&\int d^Dz_1\ldots \int d^D z_k\sum_{j=1}^k\alpha_j\Delta(x-z_j)\exp\Big(\sum_{l>j}^k\alpha_j\alpha_l\Delta(z_j-z_l)\Big)|_{\alpha^{(k)}=0} \ .
\label{unpunto}\end{align}
Expanding the exponential one sees that the total degree in the $\alpha_j$'s is odd, so that, as obvious, $\langle\phi(x)\rangle=0$ for $n$ even.
More generally, when $kn$ is even, there are no terms in (\ref{unpunto}) containing $\alpha_1^n\cdots\alpha_k^n$. Therefore, the contributions to
$\langle\phi(x)\rangle$ arise only for $kn$ odd. In this case the term
including $\alpha_1^n\cdots\alpha_k^n$ in the integrand of (\ref{unpunto}) is
\be
{1\over ({kn-1\over2})!}\sum_{j=1}^k\alpha_j \Delta(x-z_j)\Big(\sum_{l>j}^k\alpha_j\alpha_l \Delta(z_j-z_l)\Big)^{kn-1\over2} \ .
\label{riferimento}\ee
Let us focus on the term
\be
\Big(\sum_{l>j}^k\alpha_j\alpha_l \Delta(z_j-z_l)\Big)^{kn-1\over2} \ .
\label{alfain}\ee
By (\ref{seidieci}), with $kn/2$ replaced by $(kn-1)/2$, one sees that, for each $l=1,\ldots,k$,
the total exponent of $\alpha_l$ in (\ref{alfain}) is
\be
p_l:=\sum_{i=1}^{l-1}m_{il}+\sum_{j=l+1}^k m_{lj} \ ,
\label{pelletre}\ee
with the $m_{jl}$'s constrained by the condition
\be
\sum_{j>l}^km_{jl}=(kn-1)/2 \ ,
\label{doppiostar}\ee
so that
\be
\sum_{l=1}^kp_l=2\sum_{l>j}^k m_{jl}=kn-1  \ .
\label{aritroso}\ee
By (\ref{riferimento}), it follows that the terms in (\ref{alfain}) contributing to $\langle\phi(x)\rangle$ are the ones containing
$\alpha_1^n\cdots \alpha_i^{n-1}\cdots\alpha_k^n$, $i=1,\ldots k$, multiplied by the $\alpha_i$ in  $\sum_{j=1}^k\alpha_j \Delta(x-z_j)$. This implies
the condition
\be
p_{l\neq i}=n \ , \qquad p_i=n-1 \ ,
\ee
$l=1,\ldots,k$, that, by (\ref{aritroso}), implies the condition on the $m_{jl}$'s (\ref{doppiostar}). Note that
inspection of (\ref{unpunto}) shows that the term $k=1$ of the series is nonvanishing only when $n=1$. Recalling then
that nonvanishing contributions to (\ref{unpunto}) arise only for $k$ odd, one gets
\begin{align}
\langle\phi(x)\rangle = & -{1\over  W^{(n)} [0]}\Big[\delta_{n,1}\lambda\langle \Delta(x-z_1)\rangle_{z_1}+\sum_{k=1}^\infty{\lambda^{2k+1}\over (2k+1)!}
\sum_{i=1}^{2k+1}\sum_{p_i=n-1,p_1=\ldots=\check p_i=\ldots=p_{2k+1}=n} \cr
& {1\over \prod_{l>j}^{2k+1}m_{jl}!}\langle \Delta(x-z_i)\prod_{l>j}^{2k+1} \Delta(z_j-z_l)^{m_{jl}}\rangle_{z_1\ldots z_{2k+1}}\Big] \ .
\end{align}
Next, note that the sum over $i$ is just the sum of $2k+1$ identical quantities. Therefore, it corresponds to $2k+1$ times an arbitrary element of the sum. We choose the one with $i=2k+1$
\begin{align}
\langle\phi(x)\rangle = & -{\lambda\over  W^{(n)} [0]}\Big[\delta_{n,1}\langle \Delta(x-z_1)\rangle_{z_1}+\sum_{k=1}^\infty{\lambda^{2k}\over (2k)!}\sum_{p_{2k+1}=n-1,p_1=\ldots=p_{2k}=n} \cr
& {1\over \prod_{l>j}^{2k+1}m_{jl}!}\langle \Delta(x-z_{2k+1})\prod_{l>j}^{2k+1} \Delta(z_j-z_l)^{m_{jl}}\rangle_{z_1\ldots z_{2k+1}}\Big] \ .
\end{align}
Let us consider the case $n=1$. Since $p_{2k+1}=0$, it follows that $m_{l 2k+1}=0$, $l=1,\ldots, 2k$. Furthermore, the
dependence on $z_{2k+1}$ only appears in $\Delta(x-z_{2k+1})$, so that, according to (\ref{usingagain}), integration over $z_{2k+1}$ gives a factor $1/m^2$. Hence, for $n=1$,
\be
\langle\phi(x)\rangle=-{\lambda\over m^2 W^{(1)} [0]}\sum_{k=0}^\infty{\lambda^{2k}\over (2k)!}\sum_{p_1=\ldots=p_{2k}=1}{1\over \prod_{l>j}^{2k}m_{jl}!} \langle \prod_{l>j}^{2k} \Delta(z_j-z_l)^{m_{jl}}\rangle \ ,
\label{series}\ee
where it is understood that the term $k=0$ in the summation is 1.
On the other hand, by (\ref{centraldue}) the expression of $W^{(n)}[0]$ for $n$ odd is an expansion on even $k$, and this coincides, for $n=1$, with the series (\ref{series}), so that
\be
\langle\phi(x)\rangle = -{\lambda\over m^2} \ .
\label{minimo}\ee
This result can be seen as a check. Actually, for $n=1$
 the theory is free, so that $\langle\phi(x)\rangle$ corresponds to the
value of $\phi$ that minimizes $m^2\phi^2/2+\lambda\phi$, that is (\ref{minimo}).

\noindent We conclude this section by considering the connected two-point function
\be
\langle\phi(x_1)\phi(x_2)\rangle={\delta^2 Z^{(n)}[J]\over \delta J(x_1)\delta J(x_2)} \ ,
\ee
where $Z^{(n)}=-\ln W^{(n)}$. We have
\begin{align}
&\langle\phi(x_1)\phi(x_2)\rangle=-\langle\phi(x_1)\rangle\langle\phi(x_2)\rangle + \Delta(x_1-x_2)+
{1\over  W^{(n)} [0]}\sum_{k=1}^\infty{(-\lambda)^k\over (n!)^k k!}\partial_{\alpha_1}^n\ldots\partial_{\alpha_k}^n \cr
&\int d^Dz_1\ldots \int d^D z_k\sum_{j,l=1}^k\alpha_j\alpha_l\Delta(x_1-z_j)\Delta(x_2-z_l)\exp\Big(\sum_{l>j}^k\alpha_j\alpha_l\Delta(z_j-z_l)\Big)|_{\alpha^{(k)}=0} \ . \cr
\label{duepunti}\end{align}
Expanding the exponential one sees that the only contributions to the series are for $kn$ even.
The term including $\alpha_1^n\cdots\alpha_k^n$ in the integrand is
\be
{1\over ({kn-2\over2})!}\sum_{j,l=1}^k\alpha_j\alpha_l \Delta(x_1-z_j)\Delta(x_2-z_l)\Big(\sum_{l>j}^k\alpha_j\alpha_l \Delta(z_j-z_l)\Big)^{kn-2\over2} \ .
\ee
Because the second summation starts from $k=2$, it follows that the $k=1$ term is not vanishing only for $n=2$. In this case the contribution to (\ref{duepunti}) is
\be
\delta_{n,2}{\lambda\over  W^{(n)} [0]} \langle\Delta(x_1-z_{1})\Delta(x_2-z_{1})\rangle_{z_1} \ .
\ee
The contribution to $\langle\phi(x_1)\phi(x_2)\rangle$ for $k=2$ is immediate
\begin{align}
{\lambda^2\over 2 n!^2(n-1)!}&2\langle\partial_{\alpha_1}^{n-1}\partial_{\alpha_2}^{n-1} \alpha_1\alpha_2 \Delta(x_1-z_1)\Delta(x_2-z_2)\Big(\alpha_1\alpha_2 \Delta(z_1-z_2)\Big)^{n-1}\rangle_{z_1z_2}|_{\alpha_1=\alpha_2=0} \cr
&
={\lambda^2\over(n-1)!}\langle\Delta(x_1-z_1)\Delta(x_2-z_2)\Delta^{n-1}(z_1-z_2)\rangle_{z_1z_2} \ .
\end{align}
Let us now consider the case $k\geq3$. By (\ref{seidieci}), with $kn/2$ replaced by $(kn-2)/2$, one sees
that the total exponent of $\alpha_l$, $l=1,\ldots k$, in
\be
\Big(\sum_{l>j}^k\alpha_j\alpha_l \Delta(z_j-z_l)\Big)^{kn-2\over2} \ ,
\label{lapotenza}\ee
is again (\ref{pelletre}), where now $0\leq m_{jl}\leq(kn-2)/2$.
The terms in (\ref{lapotenza})
contributing to $\langle\phi(x_1)\phi(x_2)\rangle$ are the ones containing
\be
\alpha_1^n\cdots \alpha_{i_1}^{n-1}\cdots\alpha_{i_2}^{n-1}\cdots\alpha_k^n \ ,
\label{primo}\ee $i_1,i_2=1,\ldots k$,
and
\be
\alpha_1^n\cdots \alpha_{i}^{n-2}\cdots\alpha_k^n \ ,
\label{secondo}\ee
$i=1,\ldots k$.
The products (\ref{primo}) and (\ref{secondo}) correspond  to $\alpha_1^n\cdots\alpha_k^n$ once multiplied by
$\alpha_{i_1}\alpha_{i_2}$ and $\alpha_{i}^2$ in $\sum_{j,l=1}^k\alpha_j\alpha_l \Delta(x-z_j)\Delta(x-z_l)$, respectively. This implies
\be
\sum_{l=1}^kp_l=2\sum_{l>j}^k m_{jl}=kn-2  \ .
\ee
The above analysis yields
\begin{align}
\langle\phi(x_1) & \phi(x_2)\rangle=-\langle\phi(x_1)\rangle\langle\phi(x_2)\rangle + \Delta(x_1-x_2)+
{1\over  W^{(n)} [0]}\Big[\delta_{n,2}\lambda \langle\Delta(x_1-z_{1})\Delta(x_2-z_{1})\rangle_{z_1}\cr
&+{\lambda^2\over(n-1)!}\langle\Delta(x_1-z_1)\Delta(x_2-z_2)\Delta^{n-1}(z_1-z_2)\rangle_{z_1z_2}\cr
&+\sum_{k=3}^\infty{(-\lambda)^k\over k!}\sum_{i_1,i_2=1}^k\sum_{\{p\}_{n,i_1,i_2}}[k|m]\langle\Delta(x_1-z_{i_1})\Delta(x_2-z_{i_2})\prod_{l>j}^{k}\Delta(z_j-z_l)^{m_{jl}}\rangle_{z_1\ldots z_{k}}\Big] \ ,
\label{duepuntinotevole}\end{align}
where, as in (\ref{edh}), if $kn$ is even, then $[k|m]:={1\over \prod_{l>j}^km_{jl}!}$,
and $[k|m]=0$ otherwise.
$\sum_{\{p\}_{n,i_1,i_2}}$ denotes the sum over all possible values of $m_{12}, m_{13},\ldots,m_{1k},m_{21},\ldots,m_{k-1 k}$, ranging between $0$ and $(kn-2)/2$, such that
\be
p_{i_1}=n-1-\delta_{i_1i_2} \ , \quad p_{i_2}=n-1+\delta_{i_1i_2} \ , \quad p_j=n \ , \, j\ne i_1, i_2  \ .
\ee
A basic check of the above construction concerns the connected two-point function of ${\lambda\over 4!}:\phi^4:$. In the standard formulation of
${\lambda\over 4!}\phi^4$, the contribution up to order $\lambda^2$ reads
\begin{align}
\langle\phi(x_1)\phi(x_2)\rangle&=\Delta(x_1-x_2)-{\lambda\over2}\langle\Delta(x_1-z)\Delta(x_2-z)\Delta(z-z)\rangle_z \cr
&+{\lambda^2\over6}\langle\Delta(x_1-z_1)\Delta(x_2-z_2)\Delta^3(z_1-z_2)\rangle_{z_1z_2} \cr
&+{\lambda^2\over4}\langle\Delta(x_1-z_1)\Delta(z_1-x_2)\Delta^2(z_1-z_2)\Delta(z_2-z_2)\rangle_{z_1z_2} \cr
&+{\lambda^2\over4}\langle\Delta(x_1-z_1)\Delta(x_2-z_2)\Delta(z_1-z_1)\Delta(z_1-z_2)\Delta(z_2-z_2)\rangle_{z_1z_2} \ .
\end{align}
Equation (\ref{duepuntinotevole}), which gives the full expansion, reproduces such a result eliminating, by construction, the $\Delta(0)$-terms, that is,
\be
\langle\phi(x_1)\phi(x_2)\rangle=\Delta(x_1-x_2)+
{\lambda^2\over6}\langle\Delta(x_1-z_1)\Delta(x_2-z_2)\Delta^3(z_1-z_2)\rangle_{z_1z_2} \ .
\ee

\section{Conclusions}\label{sec-conc}

Let us conclude by shortly summarizing the main results in the paper.
We started by observing that the translation operator $\exp(\alpha{\delta_{J(x)}})$
can be used to get the generating functional associated to exponential interactions.
Then, it has been shown that such a translation operator generates the normal ordering terms. The analysis naturally leads to
an alternative representation of the generating functional.

\vspace{0.1cm}

\noindent
Next, we derived the scaling relations coming from the absorption of the normal ordering
contributions, and discussed the massless case, suggesting a modified Feynman propagator
(\ref{further}), whose dependence on the free parameter $m_0$ should be compared with the
ambiguities of taking the $D\to 4^+$ limit. The scaling relations of the mass are related to
the scaling properties of the generating functionals, reported in Eq.(\ref{gf}).
This follows by the integrated Schwinger-Dyson equation and the equation relating the derivative of $W[J]$ with respect to the
scale $\mu_0$ to a shift of $\phi$ in the term $\int d^D x J(x)\phi(x)$.

\vspace{0.1cm}

\noindent
We then proposed a new approach to quantum field perturbation theory: instead of perturbating the
free theory, one may use $:\exp(\alpha\phi):$ as master potential. This leads to the explicit expression
of the generating functional, reported in Eq.(\ref{centraltre}).  As a result, the
 $\Delta(0)$-terms associated to non-normal ordered potentials are absorbed at once. Furthermore,
the functional derivatives with respect to $J$, coming from the action of the potential on the free generating functional,
are replaced by ordinary derivatives with respect to auxiliary parameters.

\vspace{0.1cm}

\noindent  We note that the present analysis, which can be extended to other theories, suggests that exponential interactions
may shed light on quantum field theories. For example, one should investigate whether the auxiliary parameters, introduced as tools to compute $W^{(n)}[J]$, may have some
interpretation. Furthermore, while here the analysis has been mainly focused on perturbative techniques, with the exponential interaction acting on the free vacuum,
one should investigate the analogous structure by considering the exact vacuum
of the interacting theory. In this respect, we note that some evidence on the triviality of the exponential interaction in $D=4$ would suggest that the exact vacuum is
in some way equivalent to the free one. This suggests that we further investigate the proposed method to get nonperturbative insigths on other theories, including $\phi^4$ in $D=4$.

\section*{Acknowledgements} It is a pleasure to thank Franco Strocchi for enlightening comments and suggestions. I also
thank Antonio Bassetto, Kurt Lechner, Pieralberto Marchetti, Paolo Pasti,
 Dima Sorokin, Mario Tonin and Roberto Volpato for interesting
discussions.

\end{document}